\documentclass[prb,twocolumn,aps,showpacs]{revtex4}
\usepackage{graphicx}
\usepackage{bm}
\usepackage{amsmath,amssymb}
\usepackage{subfigure}
\usepackage{float}
\usepackage{latexsym}
\usepackage{epstopdf}
\begin{document}

\title{Interplay of spin-orbit coupling and Zeeman splitting in the absorption lineshape of 2D fermions}
\author{R. Glenn, O. A. Starykh, and M. E. Raikh}
\affiliation{Department of Physics, University of Utah, Salt Lake City, UT 84112}

\begin{abstract}
We suggest  that electron spin resonance  (ESR) experiment can be used as
a probe of spinon excitations of hypothetical spin-liquid state of
frustrated antiferromagnet in the presence of asymmetric Dzyaloshinskii-Moriya (DM) interaction.
We describe assumptions under which the ESR response
is reduced to the response of 2D electron gas with Rashba spin-orbit
coupling.
Unlike previous treatments, the spin-orbit coupling, $\Delta_{SO}$,
is not assumed small compared to the Zeeman splitting, $\Delta_Z$.
We demonstrate that
ESR response diverges at the edges of the absorption spectrum
for ac magnetic field perpendicular to the static field.
At the compensation point,
$\Delta_{SO}\approx \Delta_Z$,
the broad absorption
spectrum exhibits features that evolve with temperature, $T$, even when $T$ is comparable to the Fermi energy.

\end{abstract}
\maketitle

\section{Introduction}
In a spin system with isotropic exchange the absorption spectrum of
ac magnetic field is a $\delta$-peak at $\omega=\Delta_Z$, where $\Delta_Z$ is the Zeeman
splitting, independently of the exchange interaction strength\cite{oshikawa2002}.
Similarly to the Kohn theorem for cyclotron resonance\cite{kohn}, this fact is the consequence
of  coupling of spatially homogeneous exciting field to the center of mass of the system.
Therefore, any deviation of the absorption spectrum from the $\delta$-shape implies a
violation of the spin-rotation symmetry which is either due to anisotropic terms in the Hamiltonian
or due to development of spontaneous ordering below critical temperature.

The subject of the present paper is the shape of the absorption in the spin-liquid
ground state \cite{balents2010} of the frustrated antiferromagnet in the presence of  Dzyaloshinskii-Moriya (DM) interaction which originates
from the spin-orbit interaction in the underlying system of electrons \cite{dzyal,moriya}.
We find that electron spin resonance (ESR), which measures the ac absorption, can be used to
probe the fractionalized spinon excitations. Specifically, we assume that the ground state of
the frustrated spin system is described by the Fermi sea of the neutral spin excitations (spinons).
Emergence of the Fermi sea of neutral excitations is at the core of the RVB state proposed
by Anderson 25 years ago \cite{rvb,pomeranchuk}.  Recently the spin liquid state with spinon Fermi surface
was  found for half-filled Hubbard model on triangular lattice \cite{motrunich2005,lee2005}
for intermediate $U/t$ values.

\begin{figure}[t]
\includegraphics[width=80mm]{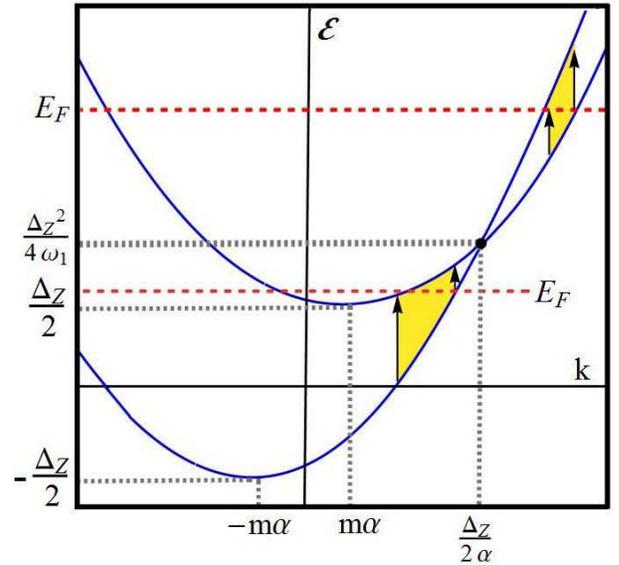}
\caption{Energy dispersion of two subbands in the presence of Zeeman splitting and spin-orbit coupling are plotted from Eq. (\ref{spectrum}). The curves cross at $k_x=\Delta_Z/2 \alpha$. Yellow regions designate the momenta of the states taking part in absorption. For Fermi-level position $E_F=\Delta_Z^2/4\omega_1$, where $\omega_1=2m\alpha^2$, there is no threshold for absorption.}
\label{figoverall}
\end{figure}
Dzyaloshinskii-Moriya   interaction that gives a nontrivial shape to the spin resonance
is most conveniently introduced into the Hubbard model via spin-dependent hopping term
\begin{eqnarray}
{\hat H} &=& \sum_{i,j} \{c^+_{i,\alpha}( - t \delta_{\alpha\beta} +
i \vec{\lambda}_{ij} \cdot \vec{\hat{s}}_{\alpha\beta})c_{j,\beta} + {\text{h.c.}}\}\nonumber\\
&&+ U \sum_i n_{i\uparrow} n_{i\downarrow} .
\label{t-dm}
\end{eqnarray}
Here 2$\vec{\hat{s}}$ is the vector of Pauli matrices, $\alpha$ and $\beta$ are spin indices, and
$\vec{\lambda}_{ij}$ is the DM vector on the link $(ij)$.
The first $(t)$ and the last $(U)$ terms are responsible for
the Fermi surface formation. Here we argue that, in course of
reduction of these terms to the Hamiltonian of spinons, as
outlined below, the second term in Eq. (\ref{t-dm}) generates
a term in the spinon Hamiltonian which has a form of spin-orbit
coupling familiar from 2D electron gas in semiconductor structures
\cite{bychkov}.

In order to illuminate the central idea of our work and to keep technical complications to a minimum,
we will  focus on the case of a {\em uniform} spin-orbit
interaction $\vec{\lambda}_{i, j} = \lambda \hat{n} \times (\vec{r}_i - \vec{r}_j)$.
Here vectors ${\vec r}_{i,j}$ denote nearest-neighbor lattice sites and $\hat{n}=\hat{z}$ is a normal to the plane.
Note in passing that, for a square lattice, precisely this arrangement is known to realize in, e.g.,  YBa$_2$Cu$_3$O$_{6+x}$
\cite{coffey1991,shekhtman1992,bonesteel1993}. For the chosen form of spin-orbit interaction the DM term
in Eq. \eqref{t-dm} assumes the form
\begin{equation}
\hat{H}_{SO}({\bf k}) = -2\lambda \sum_k c^\dagger_{k,\alpha} \{ \hat{s}_x \sin[k_y] - \hat{s}_y \sin[k_x]\} c_{k,\beta} .
\label{so-lattice}
\end{equation}
One can recognize in Eq. (\ref{so-lattice}) the lattice version of the celebrated Rashba spin-orbit term\cite{rashba60,bychkov} in the Hamiltonian
of a 2D electron gas.

Technically, the reduction of the full electronic Hamiltonian, Eqs. \eqref{t-dm} and \eqref{so-lattice},
to that describing spinon excitations of the uniform spin-liquid state \cite{lee2005}
is achieved with the help of  the slave-rotor technique
\cite{georges2004,param2007,podolsky2009}. Within this technique, a spinon $f_{i, \alpha}$ with spin projection $\alpha$ at $i$-th site
inherits the spin of the electron, $c_{i, \alpha}$, while the rotor
variable, $e^{i\theta_i}$, inherits its charge,
$c_{i, \alpha} = f_{i, \alpha} e^{-i\theta_i}$. Following the approach initially outlined in Ref. \onlinecite{georges2004},
one then substitutes this parametrization into the original Hamiltonian Eq. \eqref{t-dm} and {\em decouples} the
resulting coupled dynamics of spinons and rotors into a sum of two separate Hamiltonians
describing spinons and rotors independently. The parameters of these Hamiltonians are determined
self-consistently and are renormalized with respect to their bare values. In particular, one finds
(see Sect. III B 1 of Ref. (\onlinecite{georges2004})) that spinons are described by the free fermion Hamiltonian
\begin{eqnarray}
{\hat H}_f &=& \sum_{i,j} \{ f^+_{i,\alpha} ( - t^{\rm eff} \delta_{\alpha\beta} + i \vec{\lambda}_{ij}^{\rm eff} \cdot \vec{\hat{s}}_{\alpha\beta})f_{j,\beta} \nonumber\\
&& - g \mu_B \vec{h}\cdot f^+_{i,\alpha} \vec{\hat{s}}_{\alpha\beta} f_{i,\beta}\},
\label{H-f}
\end{eqnarray}
the parameters of which -- $t^{\rm eff}, \vec{\lambda}^{\rm eff}$ -- depend on the rotor Hamiltonian, ${\hat H}_\theta$,
the precise form of which is not needed here
(it is given by Eq. \eqref{TempEDSR} of Ref. (\onlinecite{georges2004})). For example, $t^{\rm eff}_{i,j} = t_{i,j} \langle \cos[\theta_i - \theta_j]\rangle_\theta$, where
the average is taken with respect to ${\hat H}_\theta$. Since $\vec{\lambda}_{i,j}$ represents the spin-dependent part of the
hopping integral, $t_{i,j}$, we expect it to renormalize similarly to the latter quantity. Note in passing that
in many cases spin-orbit interaction can be gauged away altogether by an appropriate unitary rotation, see Ref. \onlinecite{shekhtman1992}.
In these cases the above statement becomes exact. Note also that the symmetry alone, namely the oddness of the spin-orbit
interaction in Eq. \eqref{t-dm} under the spatial inversion, dictates that the spin-orbit interaction of spinons has the
same functional form.

The last term in Eq. \eqref{H-f} represents Zeeman coupling of electron's
spin to the external magnetic field, $\vec{h}$, which does not involve charge degrees of freedom at all, and thus
preserves its form for spinons as well.
In general, magnetic field also couples to the orbital motion of spinons via the higher order closed loop
processes\cite{motrunich2006} ($\propto t^3/U^2$ for triangular lattice, for example).
For the sake of argument we do not consider these processes here: for example ``orbital" coupling is absent in geometry where magnetic field is parallel to the two-dimensional layer.

 We would like to emphasize that the Hubbard interaction, $U$, does not enter Eq. \eqref{H-f} at all. In the slave-rotor representation it is present only in the rotor Hamiltonian ${\hat H}_\theta$
and affects spinons indirectly, via renormalized parameters such as $t^{\rm eff}$ in Eq. \eqref{H-f}, see Refs. \onlinecite{georges2004,param2007,podolsky2009}.
This feature of the slave-rotor theory makes it clear that, as long as the spin-orbit and Zeeman
interactions are small perturbations to electron's Hamiltonian (which, practically speaking, is always the case),
the spinon sector of the Hubbard model must be described by Eq. \eqref{H-f}, so that the spatial structure
of the spinon spin-orbit coupling, $\vec{\lambda}^{\rm eff}_{i,j}$, inherits that of the electron
spin-orbit term in Eq. \eqref{t-dm}. Here it means that spin-orbit part of Eq. \eqref{H-f} is given in the momentum space by
 Eq. \eqref{so-lattice} when electron operators $c_{k,\alpha}$ are replaced by spinon $f_{k,\alpha}$ ones.

Finally, the spin-liquid phase we are interested in is just a disordered (Mott) phase of ${\hat H}_\theta$,
in which the expectation value of the rotor field is zero, $\langle e^{i \theta}\rangle = 0$,
and the charge degrees of freedom are gapped.
(The phase with a condensate of charge rotor field $\langle e^{i \theta}\rangle \neq 0$
describes the usual metallic phase of Eq. \eqref{t-dm} and is not of interest here.)
Importantly, spin excitations of this phase
are neutral spinons, $f_{i, \alpha}$, forming the Fermi-liquid state and
described by the Hamiltonian Eq. \eqref{H-f} with parameters which are
renormalized by the gapped charge fluctuations
(the notion that spinons can be treated as neutral fermions  \cite{motrunich2005,lee2005,mross2011,zhou2011}
has an important caveat that they interact strongly with emerging gauge field \cite{palee2006,podolsky2009}).

%

We thus see that, under the assumptions described above, the Hamiltonian of strongly interacting
Fermi system Eq. \eqref{t-dm} in the spin-liquid phase 
maps onto that of  non-interacting spinon Fermi gas with spin-orbit interaction of Rashba type \eqref{H-f}.
Hence, the basic features of the ESR response in an exotic spin liquid state
can be captured within a much simpler model of two-dimensional electron gas
subject to spin-orbit interaction.
Surprisingly, despite several decades of intensive research
in the latter area, we are not aware of the calculation of ESR response in two-dimensional
electron gas subject to both spin-orbit and Zeeman interactions.

\begin{figure}[h]
\includegraphics[width=70mm]{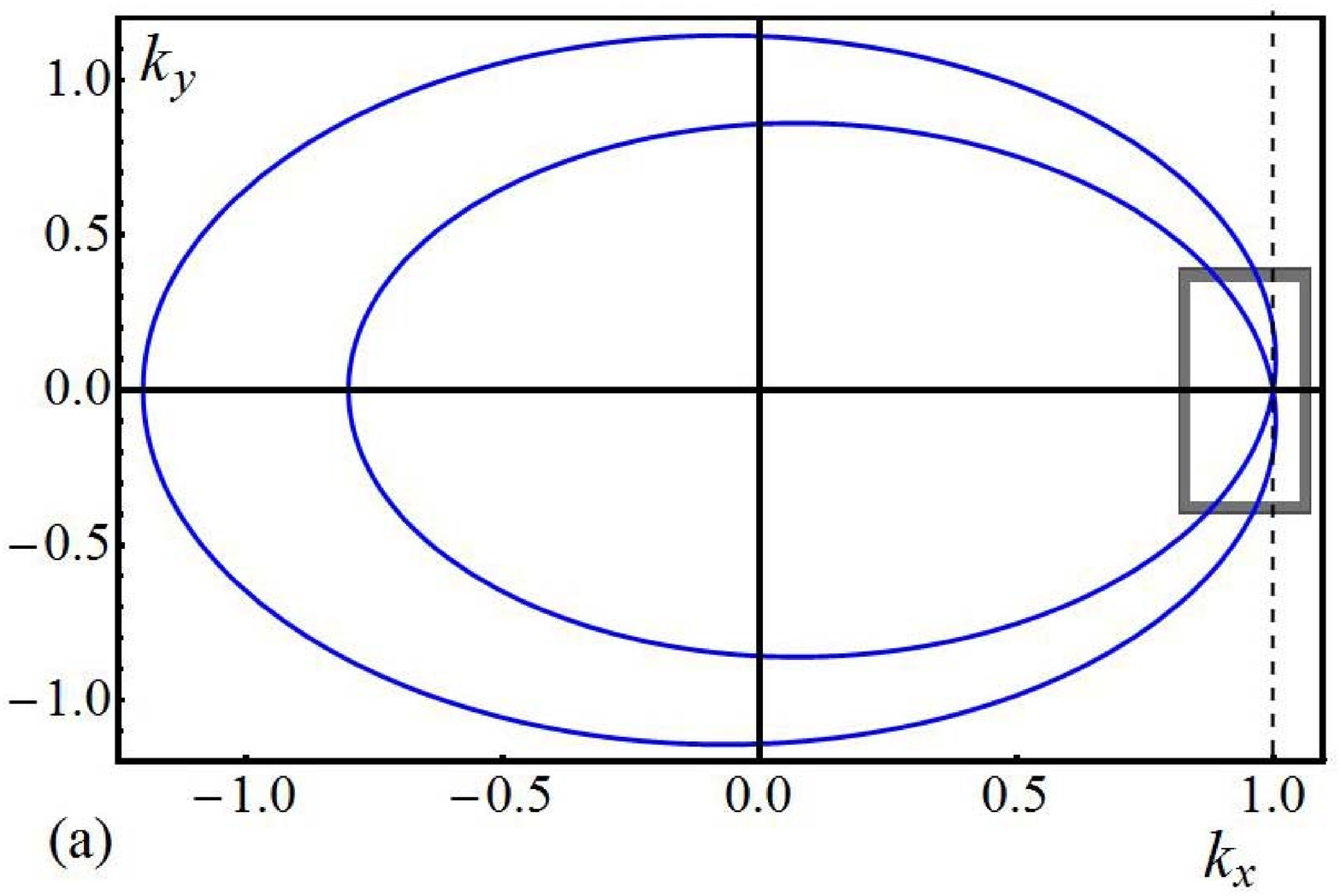}
\includegraphics[width=70mm]{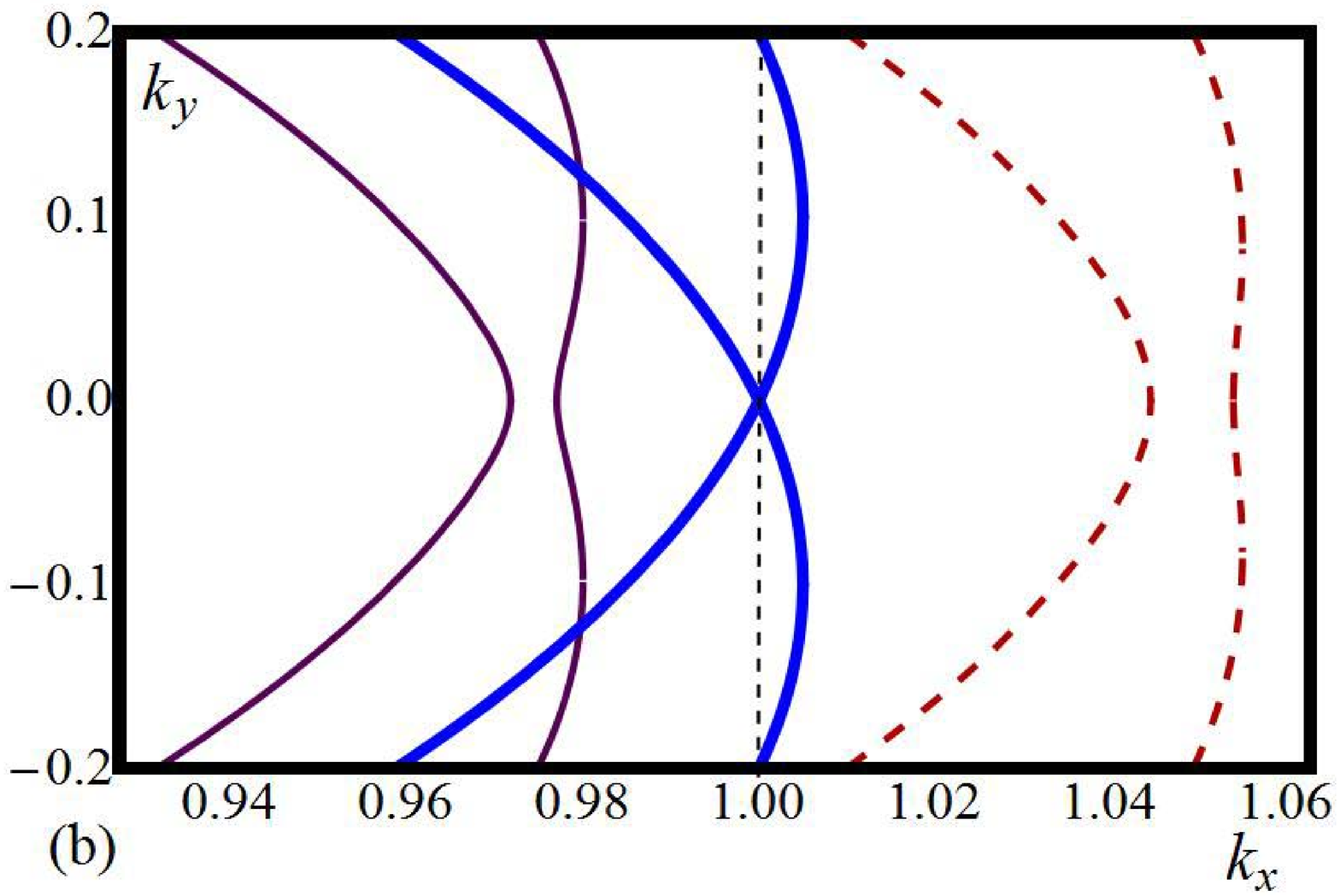}
\vspace{-0.5cm}
\caption{(a) Two Fermi surfaces corresponding to $E_F^{(0)}= \Delta_Z^2/ 4\omega_1$ and $\omega_1/\Delta_Z=0.1$ are plotted from Eq. (\ref{fermisurface}), where the projections of momenta $k_x$ and $k_y$ are in the units $\Delta_Z/2 \alpha$.
(b) Evolution of the Fermi surfaces near the point $k_y=0$ for different $E_F$ (enclosed by the gray box in (a)): (thin line) $E_F=0.95E_F^{(0)}$, (thick line) $E_F=E_F^{(0)}$, (dashed line) $E_F=1.1E_F^{(0)}$.}
\label{figmomentum}
\end{figure}

In earlier studies of spin-orbit coupled two-dimensional electron gas the
combined effect of Zeeman field and spin-orbit interaction
was considered in relation to electric dipole spin resonance\cite{rashba63}  (EDSR),
which is the absorption of the ac electric field between the Zeeman-split levels.
Obviously, it is spin-orbit interaction which allows this absorption to occur,
since it couples the electron spin to the electric field. In theoretical papers\cite{rashba03,Efros3} on EDSR in two-dimensional
electron gas a spin-orbit term of the form Eq. (\ref{so-lattice}) in the small-${\bf k}$ limit was added to
the Hamiltonian of free electrons. Then the matrix element of transition between two spin levels was calculated for different orientations of magnetic field.  However, we cannot use the results of Refs.~\onlinecite{rashba03,Efros3}
for finding the lineshape of ESR. This is because in these papers it was assumed that the magnitude,
$\Delta_{SO}$, of the spin-orbit term is much smaller than the Zeeman splitting, $\Delta_Z$, so that  $\Delta_{SO}$ was treated perturbatively. As a result, in the absence of disorder, the lineshape of the resonance was simply a $\delta$-peak. The natural basis in which the magnitude of this peak is calculated \cite{rashba03,Efros3}
is the basis of spin projections on the applied magnetic field.

Another group of relevant earlier papers\cite{Finkelstein,Farid} dealt
with the phenomenon of chiral resonance, which takes place when $\Delta_{SO}$
is finite. In this case spin-orbit term splits
the free-electron spectrum into two branches,
separated by $\Delta_{SO}$, which correspond to different chiralities.  Then the dipole absorption of
the ac electric field
between the branches is allowed.  The absorption spectrum of the chiral
resonance has a box-like shape\cite{Farid} centered at $\omega=\Delta_{SO}$ with a
width of the order of $\Delta_{SO}^2/E_F$, where $E_F$ is the Fermi energy.  The absorption
is calculated in a natural basis of spinors, describing opposite chiralities.
Still, we cannot use the results of Refs. \onlinecite{Finkelstein,Farid} for calculating the ESR lineshape,
since magnetic field was assumed zero in these papers.
Instead, we will adopt the general approach of Ref. \onlinecite{Farid}
to the calculation of the ac response. This approach is based
on the calculation of the  {\em optical conductivity}.
We will subsequently demonstrate that the ESR lineshape
can be expressed through the optical conductivity in a
simple way.

The outline of the paper is as follows.
Calculation of the optical conductivity
 for arbitrary relation between $\Delta_{SO}$ and $\Delta_Z$,
is presented in Sect. II of the present paper. In Sect. III. the
results of this calculation are used to find the ESR lineshape
for different orientations of the ac magnetic field.

Normally,  $\sigma(\omega)$ is a peak
with two sharp edges, which are imposed by the energy conservation,
as illustrated in Fig. \ref{figoverall}. Notably, we demonstrate that $\sigma(\omega)$
exhibits a singular behavior near these  edges. The character of singularities is different
for ac field parallel and perpendicular to the external in-plane magnetic field.
We trace how these singularities are smeared out
upon increasing the temperature, $T$.

The most nontrivial behavior of $\sigma(\omega)$ corresponds  to the situation  $\Delta_Z=\Delta_{SO}$,
when Zeeman splitting is ``compensated" by the spin-orbit splitting along the direction of the magnetic field. As a result of this compensation, the two branches of electron spectrum, $\mathcal{E}^{(1)}(\mathbf{k})$ and $\mathcal{E}^{(2)}(\mathbf{k})$, cross each other at certain momentum $\mathbf{k}=(k_0,0)$ while the Fermi level $E_F=\mathcal{E}^{(1)}(\mathbf{k})=\mathcal{E}^{(2)}(\mathbf{k})$ is located at the point of crossing, as illustrated
in Fig. \ref{figoverall}. The underlying reason why the shape of $\sigma(\omega)$ is peculiar near the condition $\Delta_Z=\Delta_{SO}$
is the following.

As the difference $\Delta_Z-\Delta_{SO}$ changes sign, the Fermi surfaces $\mathcal{E}^{(1)}(\mathbf{k})=E_F$ and $\mathcal{E}^{(2)}(\mathbf{k})=E_F$  undergo rapid restructuring.
This restructuring is illustrated in Fig.~\ref{figmomentum}.
Since the states responsible for absorption lie between the two Fermi points, see Fig.~\ref{figoverall}, in the vicinity of $\Delta_Z=\Delta_{SO}$ the absorption remains finite at low frequencies $\omega \ll \Delta_Z$.
We will show that in this regime the absorption spectrum exhibits features that evolve with temperature,
even when $T$ is comparable to the Fermi energy.

\section{Optical conductivity with finite $\Delta_Z$ and $\Delta_{SO}$}

\subsection{Hamiltonian and eigenfunctions}
Assume that electrons are confined to the $(x,y)$ plane, while magnetic
field, creating the splitting, $\Delta_Z$,  is directed along the $y$-axis. For concreteness we choose the
Rashba-type\cite{rashba60,bychkov} form of spin-orbit coupling
\begin{equation}
\label{SOcoupling}
{\hat H}_{SO}({\bf k})=2\alpha(k_x\hat{s}_y-k_y\hat{s}_x),
\end{equation}
where $\alpha$ is the spin-orbit constant.
In the matrix form, the Hamiltonian of the system reads
\begin{eqnarray}
\label{Hamiltonian EDSR}
\!\!{\hat H}=\left(\begin{array}{cc}
\frac{\hbar^2}{2m}(k_x^2+k_y^2 )& i\frac{\Delta_Z}{2}- \alpha (i k_x+k_y)\\
\,\\
-i\frac{\Delta_Z}{2} - \alpha (-i k_x+k_y)  &\frac{\hbar^2}{2m}(k_x^2+k_y^2 ) \\
\end{array}\right )\!\!,
\end{eqnarray}
where $m$ is the effective mass (we set $\hbar=1$ throughout the paper).
The two branches of the spectrum and corresponding spinors are given by
\begin{equation}
\label{spectrum}
\mathcal{E}_{\mathbf {k}}^{1(2)}= \frac{k^2}{2m} \pm \sqrt{\alpha^2 k_y^2 + \left(\alpha k_x - \frac{\Delta_Z}{2} \right)^2},
\end{equation}
\begin{eqnarray}\label{spinor}
  u_{\mathbf{k}}^{(1),(2)} =\frac{1}{\sqrt 2}\left(\begin{array}{cc}
\,\frac{ \alpha(ik_x +k_y)-i\frac{\Delta_Z}{2}}{\frac{k^2}{2m}  - \mathcal{E}_{\mathbf {k}}^{1(2)}}\\
\,\\
1\\
\end{array}\right)\!\!.
\end{eqnarray}
To express the optical conductivity in terms of the spectrum Eq. (\ref{spectrum})
and eigenfunctions Eq. (\ref{spinor}) one has to add to the Hamiltonian
Eq. (\ref{Hamiltonian EDSR}) the term
\begin{equation}
\label{interaction}
{\hat H}_{int}^{e}= e{\bf E}\cdot {\bm{\rho}}\cos\omega t,
\end{equation}
where ${\bf E}\cos\omega t$ is the external ac electric field, and calculate
the ac current. This calculation is not straightforward because the
eigenfunctions from different branches corresponding to the {\em same}
momentum are orthogonal to each other.
For this reason, it is much more convenient to adopt the approach of
Ref. \onlinecite{Reizer}. Within this approach the system response to the scalar
perturbation
\begin{equation}
\label{electric}
V^e({\bm{\rho}},t)= V_0\cos(\omega t- {\bf q}\cdot{\bm{\rho}}),
\end{equation}
with finite wave vector, ${\bf q}$, is studied. The energy absorption rate, $I({\bf q})$, caused by this
perturbation, is calculated using the Golden rule.
The optical conductivity at finite ${\bf q}$ is related to $I({\bf q})$ as follows
\begin{equation}
\label{sigma}
\sigma({\bf q},\omega)=\frac{\big(1-e^{-\omega/T}\big)I({\bf q})}{2V_0^2 q^2},
\end{equation}
where $\exp(-\omega/T)$ accounts for the emission processes. The static limit ${\bf q}\rightarrow 0$
is taken only at the final step yielding
\begin{eqnarray}
\label{main}
\sigma(\omega)\!\!\!&=&\!\!\!\pi e^2
\big(1-e^{-\omega/T}\big )\lim_{{\bf q} \rightarrow 0} \left (\frac{\omega}{q^2}\right )\!\sum_{\bf k} \big |\langle u_\mathbf{k}^{(1)} |u_{\mathbf{k}+\mathbf{q}}^{(2)}\rangle \big |^2
\nonumber \\
 &&
\times \delta\big( \mathcal{E}_\mathbf{k+q}^{(2)} - \mathcal{E}_\mathbf{k}^{(1)} - \omega \big)\!\Big[1- f\big( \mathcal{E}_\mathbf{k+q}^{(2)} \big)\!\Big]
 f\big(\mathcal{E}_\mathbf{k}^{(1)}\big),\nonumber \\
\end{eqnarray}
where $f(\mathcal{E})=\Bigl\{1+\exp[(\mathcal{E}-E_F)/T]\Bigr\}^{-1}$ is the Fermi distribution.

The formal reason why absorption of the ac field is permitted in the presence of spin-orbit coupling is that,
at small ${\bf q}$, the product $\langle u_{\mathbf{k}}^{(1)}|u_{\mathbf{k}+\mathbf{q}}^{(2)}\rangle$
is a linear function of ${\bf q}$, as readily follows from Eq. (\ref{spinor})
 \begin{equation}
\label{sozscalarproduct}
\langle u_{\mathbf{k}}^{(1)}|u_{\mathbf{k}+\mathbf{q}}^{(2)}\rangle= i\Big (\frac{\alpha}{2}\Big )\frac{(\alpha k_x-\frac{1}{2}\Delta_Z)q_y -\alpha k_y q_x}{(\alpha k_x-\frac{1}{2}\Delta_Z)^2 +\alpha^2 k_y^2},
\end{equation}
so that the combination, $|\langle u_{\mathbf{k}}^{(1)}|u_{\mathbf{k}+\mathbf{q}}^{(2)}\rangle|^2/q^2$,
 remains finite in the limit of $\mathbf{q} \rightarrow 0$. In subsequent sections we analyze Eq. (\ref{main}) first at $T=0$,
and then at finite temperatures.
\subsection{Optical conductivity at $T=0$}
It follows from Eq. (\ref{sozscalarproduct}) that at finite Zeeman splitting the optical conductivity is
highly anisotropic.
For polarization perpendicular to the magnetic field, vector ${\bf q}$ is directed along the $x$-axis.
 Substituting Eq. (\ref{sozscalarproduct}) into Eq. (\ref{main}), and  converting the sum into an integral in
polar coordinates, we get
\begin{eqnarray}
\label{zeemanab}
\sigma_{\perp}(\omega)\!\!\!& =&\!\!\! \left ( \frac{e^2 \alpha^4 \omega}{16 \pi } \right )
\int\! dk k^3 \!\!\int \!\frac{d\phi}{\Omega(\phi)^4 }\sin^2 \phi \,\,
\delta \big(2 \Omega(\phi) -\omega \big)
\nonumber \\
&&\hspace{-0.3cm}\times \Theta \left(\frac{k^2}{2m} + \Omega(\phi) - E_F \right )
 \Theta \left(E_F -\frac{k^2}{2m} + \Omega(\phi) \right )\!\!, \nonumber \\
\end{eqnarray}
 where we have introduced the notation
\begin{equation}
\Omega(\phi) =\sqrt{ \alpha^2 k^2 -\alpha \Delta_Z k \cos \phi+ \frac{\Delta_Z^2}{4}}.
\end{equation}
The factor $\Omega(\phi)^{-4}$ in the integrand emerges from the denominator in
the  product $\langle u_{\mathbf{k}}^{(1)}|u_{\mathbf{k}+\mathbf{q}}^{(2)}\rangle$, see
Eq. (\ref{sozscalarproduct}). The numerator in this product is proportional to $k_y$, giving rise to $\sin^2\phi$
in the integrand. It is convenient to first perform the angular integration in Eq. (\ref{zeemanab}) with the help of the
$\delta$-function. Then the remaining integral over $k$  can be cast in the form
\begin{eqnarray}
\label{zeemanab2}
\hspace{-0.5cm}\sigma_{\perp}(\omega)\!\!\!&=&\!\!\!
\left ( \frac{e \,\alpha}{\sqrt{\pi}\,\omega \,\Delta_Z} \right ) ^2
\nonumber\\
&&\hspace{-1.5cm}\times\int\limits_{\sqrt{m(2E_F -\omega)}}^{\sqrt{m(2E_F +\omega)}} \hspace{-0.5cm}dk \, k
\left[ \alpha^2 k^2 \Delta_Z^2 - \left(\alpha^2 k^2 + \frac{\Delta_Z^2}{4} - \frac{\omega^2}{4} \right )^2\right]^{1/2}\hspace{-0.4cm}.
\nonumber\\
\end{eqnarray}
Integration in Eq. (\ref{zeemanab2}) can be performed explicitly, yielding
\begin{eqnarray}
\label{perp}
\sigma_{\perp}(\omega)\!\!\!&=&\!\!\!\left ( \frac{e^2}{8 \pi} \right ) \int_{r_i}^{r_f} dr\sqrt{ 1- r^2}
\nonumber \\
&&\hspace{-0.2cm}=\left ( \frac{e^2}{16 \pi} \right )\left( \arcsin r  + r\sqrt{1-r^2}\right )\Bigg|_{r_i}^{r_f}.
\end{eqnarray}
The limits of integration in Eq. (\ref{perp})  are given by
\begin{equation}
\label{r}
r _{f,i}(\omega)
=\frac{1}{2\Delta_Z \omega}( \Delta_{SO}^2-\omega^2 \pm 2\omega_1 \omega -\Delta_Z^2),
\end{equation}
where the frequency $\omega_1$ is defined as
\begin{equation}
\label{omega1}
\omega_1=2m\alpha^2=\frac{\Delta_{SO}^2}{4E_F}.
\end{equation}
Expression Eq. (\ref{perp}) is valid when both $r_i$, $r_f$ lie within interval $-1<r_i,r_f<1$.
For frequencies at which both $r_i$ and $r_f$ are outside this interval, we have $\sigma_{\perp}(\omega)~\equiv~0$.
When only one of the limits, say $r_f$, is outside the interval, one should set $r_f=1$ in Eq. (\ref{perp}).

For polarization parallel to the magnetic field, the calculation of conductivity is completely similar
to the calculation of $\sigma_{\perp}(\omega)$. The only difference is in the product,
$\langle u_{\mathbf{k}}^{(1)}|u_{\mathbf{k}+\mathbf{q}}^{(2)}\rangle$,
which, for ${\bf q}$ directed along $y$, is proportional to $(\alpha k_x - \Delta_Z/2)$.
In polar coordinates this factor leads to replacement of $\sin^2\phi$ in the integrand of
Eq. (\ref{zeemanab}) by $(\cos\phi -\Delta_Z/\alpha k)^2$.
As a result, instead of Eq. (\ref{perp}), the momentum integration reduces to

\begin{eqnarray}
\label{parallel}
\sigma_{\|}(\omega)\!\!\!&=&\!\!\!\left ( \frac{e^2}{8 \pi} \right ) \int_{r_i}^{r_f} dr\frac{r^2}{\sqrt{ 1- r^2}}
\nonumber \\
&&\hspace{-0.2cm}=\left ( \frac{e^2}{16 \pi} \right )\left( \arcsin r  - r\sqrt{1-r^2}\right )\Bigg|_{r_i}^{r_f}.
\end{eqnarray}
Equations (\ref{perp}) and (\ref{parallel}) constitute our main results. We start their
analysis from the limiting cases $\Delta_{SO}\gg \Delta_Z$ and $\Delta_{SO}\ll \Delta_Z$.

\subsubsection{$\Delta_{SO}\gg \Delta_Z$}

It is instructive to trace how the shape of the chiral resonance
emerges from Eqs. (\ref{perp}), (\ref{parallel}). Taking the limit $\Delta_Z \rightarrow 0$,
we find that $r_i$ and $r_f$ should be set to $-1$ and $1$, respectively, for $\omega$ within the interval $|\omega-\Delta_{SO}| < \omega_1$. Outside of this interval we have $r_i=r_f= \pm 1$ and the absorption is zero. This yields the lineshape
\begin{eqnarray}
\label{Farid}
\hspace{-1.4cm}\sigma(\omega)\!=\!\left ( \frac{e^2}{16 } \right )\!
\Theta\!\!\left [\! 1+\!\left (\frac{\omega^2 -\Delta_{SO}^2}{2 \omega \omega_1}\right) \right ]\!
\Theta\! \!\left [\! 1-\!\left (\frac{\omega^2 -\Delta_{SO}^2}{2 \omega \omega_1}\right) \right ]\!\!,\hspace{-1.5cm}\nonumber\\
\end{eqnarray}
with sharp ends, which can be further simplified by introducing $\delta\omega=\omega-\Delta_{SO}$ and taking
into account that $\omega_1\ll \Delta_{SO}$. Then we reproduce the result of Refs. \onlinecite{Farid, Magarill}
\begin{equation}
\label{Farid1}
\sigma( \delta\omega) = \frac{\,e^2}{16 }\, \Theta \left( 1-\frac{|\delta \omega|}{ \omega_1} \right )\!\!,
\end{equation}
where $\Theta(x)$ is the step-function.

Next we study how a small Zeeman splitting, $\Delta_Z \ll \omega_1$, smears the steps at
$\omega-\Delta_{SO}= \pm \omega_1$.  Introducing dimensionless deviation from the right boundary
\begin{equation}
\label{eta}
\eta(\omega)=\frac{(\Delta_{SO}-\omega + \omega_1)}{ \Delta_Z},
\end{equation}
we get the following behavior of $\sigma_{\perp}$, $\sigma_{\|}$ near the right boundary
\begin{equation}
\label{boundaryboth}
\sigma_{\perp,\|}=\left(\frac{e^2}{16\pi}\right)\left[ \frac{\pi}{2}-\arcsin \eta(\omega) \mp \eta(\omega) \sqrt{1-\eta^2(\omega)}\right]\!\!.
\end{equation}
\begin{figure}
\includegraphics[width=70mm]{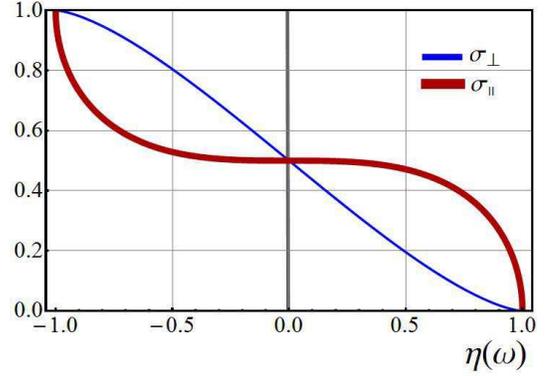}
\caption{Smearing of the right edge of the chiral resonance at small $\Delta_Z \ll \omega_1$ is plotted from Eq. (\ref{boundaryboth}) as a function of dimensionless deviation, $\eta(\omega)$, Eq. (\ref{eta}), from the boundary. $\sigma_{\|}$  and $\sigma_{\perp}$ are plotted in the units $e^2/16$.
}
\label{fig smearing}
\end{figure}
This behavior is illustrated in Fig. \ref{fig smearing}. We see that a step at $\delta\omega=\omega_1$
extends, at finite $\Delta_Z$, over an interval $\delta\omega-\omega_1=\pm \Delta_Z$. Optical conductivity is
strongly anisotropic within this interval.  In particular, $\sigma_{\|}$ has a zero slope at the center of the interval. As $\Delta_Z$ exceeds $\omega_1$, the flat top in optical conductivity disappears completely.

\subsubsection{$\Delta_Z\gg \Delta_{SO}$}

Consider now the EDSR limit  of strong Zeeman splitting, $\Delta_Z \gg \Delta_{SO}$.  Then  Eqs. (\ref{perp}), (\ref{parallel}) reduce to
\begin{equation}
\label{EDSRperp}
\sigma_{\perp}(\omega)=
\left ( \frac{\sqrt{2}e^2}{4\pi} \right )
\bigg(\frac{\omega_1}{\Delta_Z}\bigg)
 \left[ \frac{\Delta_{SO}^2-(\omega-\Delta_Z)^2}{2\Delta_Z \omega}\right]^{1/2}\!\!,
\end{equation}
\begin{equation}
\label{EDSRparallel}
\sigma_{\|}(\omega)=\left ( \frac{e^2}{4\sqrt{2}\pi} \right )
\bigg(\frac{\omega_1}{\Delta_Z}\bigg)
\left[ \frac{\Delta_{SO}^2-(\omega-\Delta_Z)^2}{2\Delta_Z \omega}\right]^{-1/2}\!\!.
\end{equation}
The fact that the EDSR spectrum is confined to the interval $|\omega-\Delta_Z| < \Delta_{SO}$ could be expected on general grounds. It is also natural that optical conductivity is proportional, via $\omega_1$, to the square of the SO coupling strength.
Less trivial is the fact that EDSR absorption is highly anisotropic, so that $\sigma_\perp$ turns to zero at the edges, while $\sigma_\|(\omega)$ exhibits a divergence near the edges.
This can be seen in Fig. \ref{figzeeman1}a.

\subsubsection{$\Delta_Z \sim \Delta_{SO}$}

Consider now the intermediate regime when $\Delta_Z$ and $\Delta_{SO}$ are of the same order. Since we assumed
that Fermi energy is bigger than both $\Delta_Z$ and $\Delta_{SO}$, the condition $\Delta_Z\sim \Delta_{SO}$ automatically implies that both $\Delta_Z$ and $\Delta_{SO}$ are much bigger than $\omega_1$.
Then the integrands in Eqs. (\ref{perp}), (\ref{parallel}) can simply be multiplied by $r_f-r_i=2\omega_1/\Delta_Z$, yielding
\begin{equation}
\label{perp1}
\sigma_{\perp}(\omega)=\left ( \frac{e^2}{8 \pi} \right )\frac{2\omega_1}{\Delta_Z} \sqrt{ 1- r^2(\omega)},
\end{equation}
\begin{figure}[t]
\includegraphics[width=70mm]{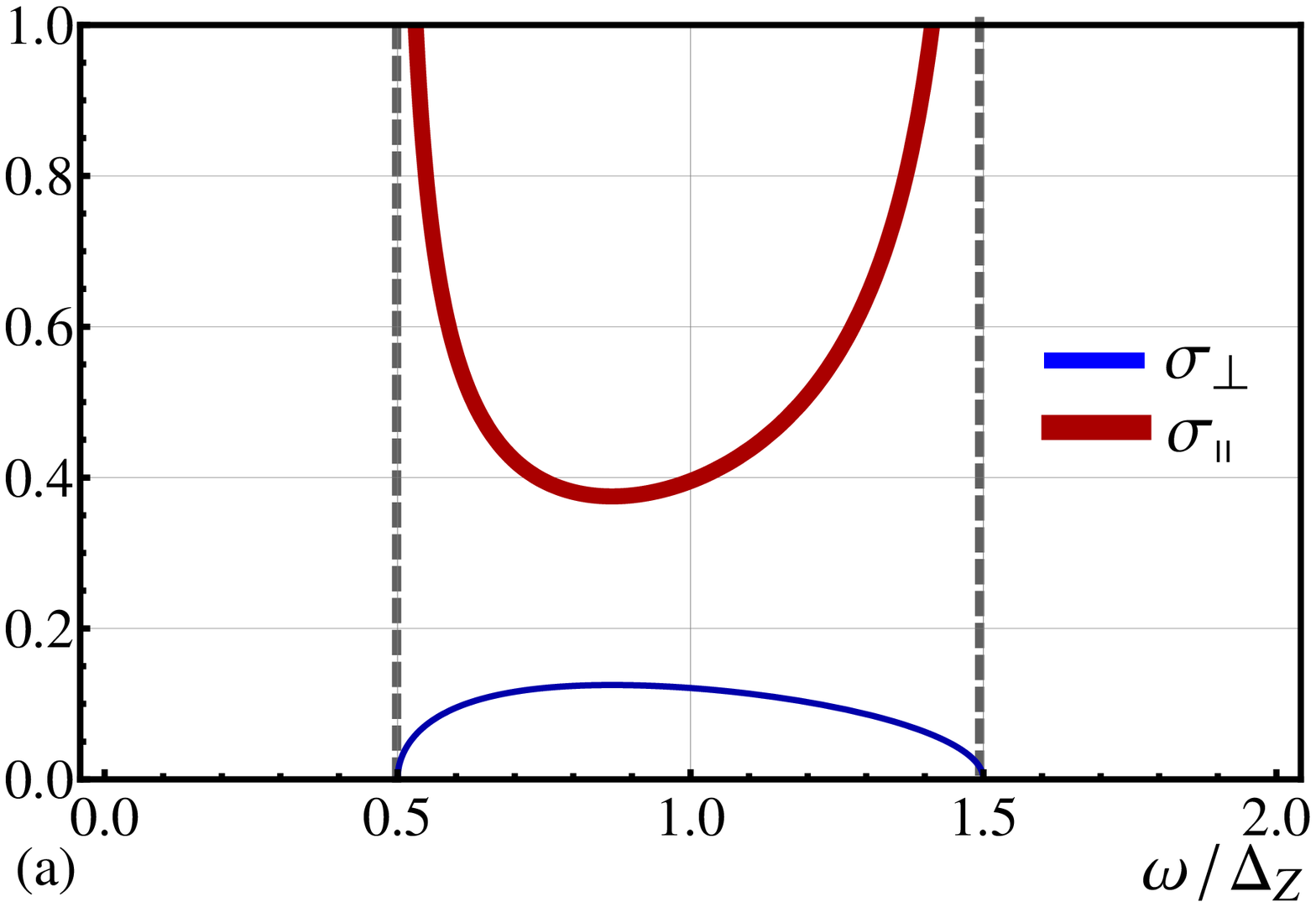}
\hspace{0.5cm}
\includegraphics[width=70mm]{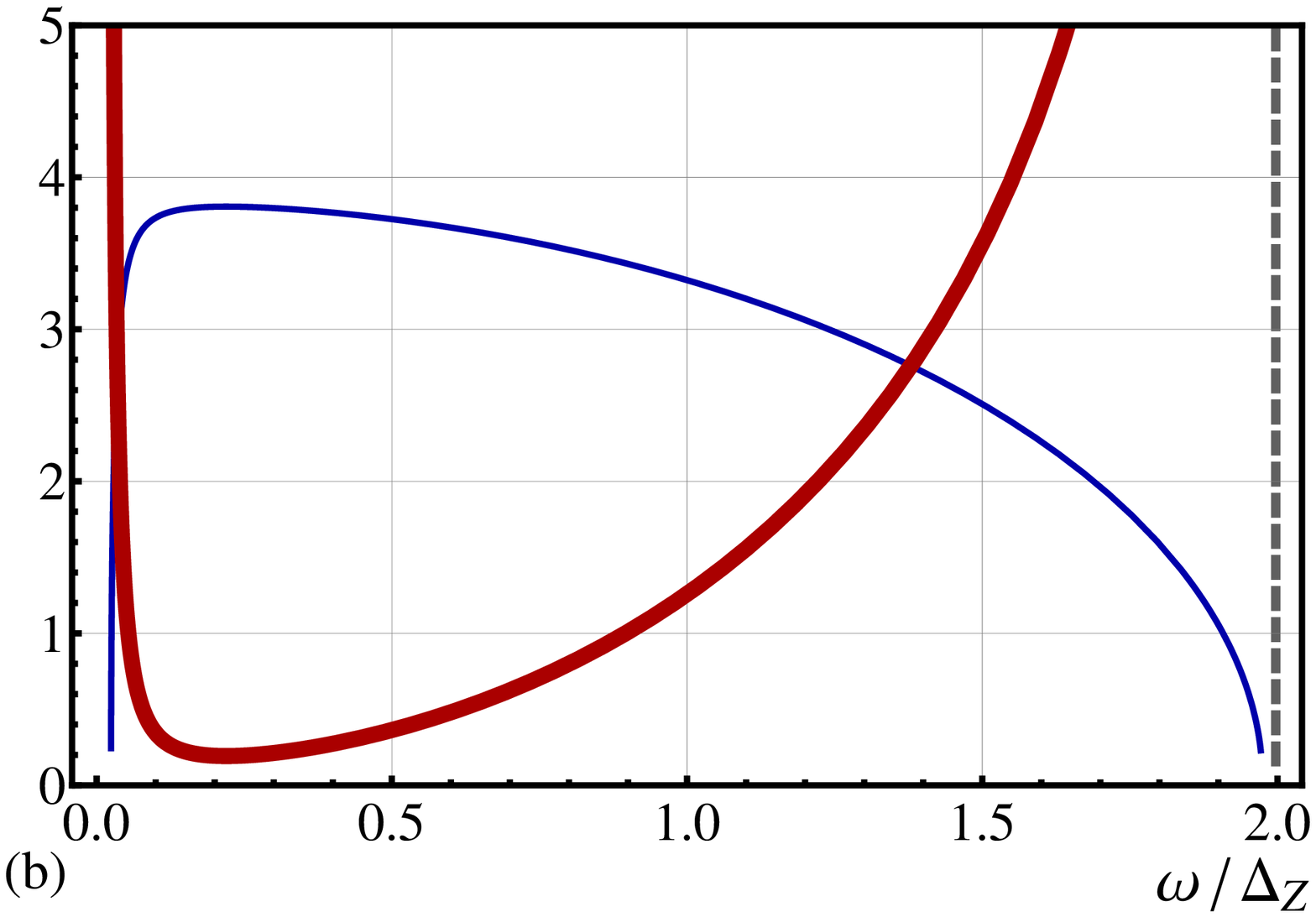}
\includegraphics[width=70mm]{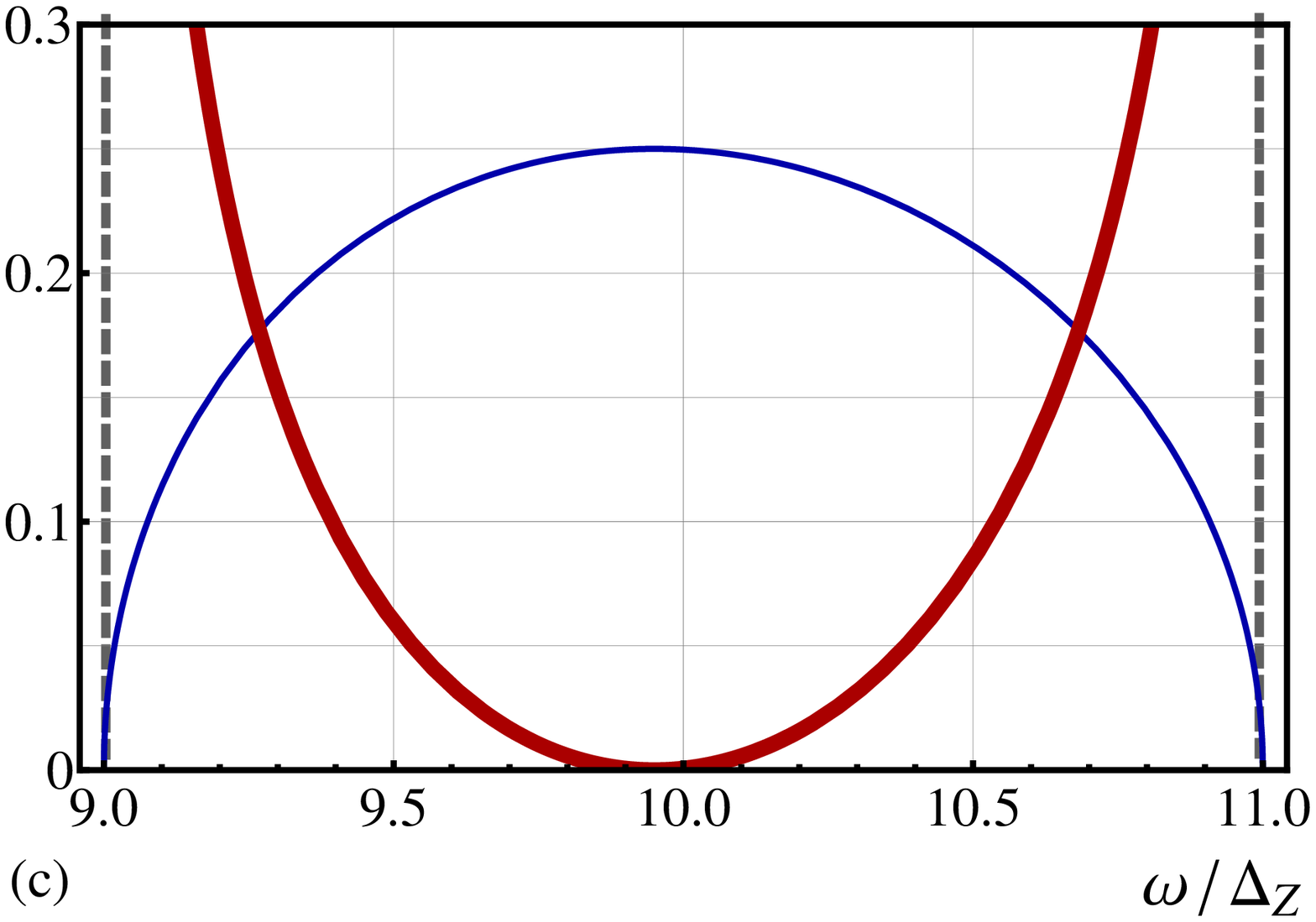}
\caption{Optical conductivity at zero temperature, Eqs. (\ref{perp1}), (\ref{parallel1}) vs. dimensionless frequency, $\omega/\Delta_Z$, are plotted in the units $ (e^2/\pi)(\omega_1/\Delta_Z)$, for three values of the ratio $\Delta_{SO}/\Delta_Z$. (a) $\Delta_{SO}/\Delta_Z=0.5$, Zeeman splitting dominates.  (b) Zeeman and SO splitting
nearly ``compensate" each other, $\Delta_{SO}/\Delta_Z=0.99$. (c) $\Delta_{SO}/\Delta_Z=10$, SO splitting dominates.
}
\label{figzeeman1}
\end{figure}
\begin{equation}
\label{parallel1}
\sigma_{\|}(\omega)
=\left ( \frac{e^2}{8 \pi} \right ) \frac{2\omega_1}{\Delta_Z}
\frac{r^2(\omega)}{\sqrt{ 1- r^2(\omega)}},
\end{equation}
where
\begin{equation}
\label{rAgain}
r(\omega)=\frac{1}{2\Delta_Z \omega}( \Delta_{SO}^2-\omega^2  -\Delta_Z^2),
\end{equation}
is given by Eq. (\ref{r}) with $\omega_1=0$. It is seen
that $\sigma_{\perp}$ monotonically grows with $r^2$, while $\sigma_{\|}$
monotonically decreases with $r^2$.
At the same time, the {\em frequency} dependencies of $\sigma_{\perp}$ and $\sigma_{\|}$
are strongly non-monotonic. Firstly, for $\Delta_{SO}>\Delta_Z$,
$r(\omega)$ turns to zero at $\omega=\omega_0=(\Delta_{SO}^2-\Delta_Z^2)^{1/2}$.
A zero in $\sigma_{\|}$ is accompanied by a maximum in $\sigma_{\perp}$. For opposite
relation, $\Delta_Z>\Delta_{SO}$, there is a minimum in $r^2(\omega)$ at $\omega=\tilde{\omega_0}=(\Delta_Z^2-\Delta_{SO}^2)^{1/2}$.
This minimum translates into a minimum in $\sigma_{\|}(\omega)$ and a maximum in $\sigma_{\perp}(\omega)$.
The behaviors of $\sigma_{\|}(\omega)$ and $\sigma_{\perp}(\omega)$ for different relations between $\Delta_Z$
and $\Delta_{SO}$ are illustrated in Fig. \ref{figzeeman1}.

The most interesting behavior of optical conductivity takes place near the point of ``compensation", where
$\Delta_Z$ and $\Delta_{SO}$ are close to each other. Then the minimum in $\sigma_{\|}$ and the maximum in $\sigma_{\perp}$ become very sharp. This implies that absorption becomes possible at very low frequencies.
On the physical level, the low-frequency absorption is the result of crossing of the Fermi surfaces $\mathcal{E}^{(1)}(\mathbf{k})=E_F$ and $\mathcal{E}^{(2)}(\mathbf{k})=E_F$, see Fig.~ \ref{figoverall}, which takes place at $\Delta_Z =\Delta_{SO}$. The Fermi energy is equal to
\begin{equation}
\label{crossing}
E_F^{(0)} = \frac{\Delta_Z^2}{ 4 \omega_1}
 \end{equation}
 at the point of crossing. As the difference  $\Delta_Z-\Delta_{SO}$ increases the Fermi surfaces undergo restructuring. In Fig. \ref{figmomentum} we show the evolution of the Fermi surfaces
\begin{equation}
\label{fermisurface}
E_F= \frac{1}{2m} \left(k_x^2 +k_y^2\right) \pm \sqrt{
\alpha^2 k_y^2 + \left( \alpha k_x -\frac{\Delta_Z}{2} \right)^2}\!,
\end{equation}
as $E_F$ departs from $E_F^{(0)}$. Since the states responsible for absorption lie between the two Fermi points, Fig.~\ref{figoverall}, the low-frequency part of the absorption spectrum follows this departure.
\subsection{Optical conductivity at  finite temperature}
\label{sec:finite-T}
We now turn to the general expression Eq. (\ref{main}) for optical conductivity
and perform integration over the directions of momentum, ${\bf k}$.   This yields
the following generalizations of Eqs. (\ref{perp}), (\ref{parallel}) to finite $T$
\begin{eqnarray}
\label{Tperpabsorp}
\sigma_{\perp}(\omega)\!\!\!&=&\!\!\!
\left ( \frac{e\, \alpha}{2\sqrt{\pi} \omega\Delta_Z} \right )^2
\sinh \left (\frac{\omega}{2T} \right )
\int_0^\infty dk \,\,k
\nonumber \\
&&\hspace{-0.3cm}\times\frac{ \sqrt{\Big[4 \alpha^2 k^2  -(\omega+\Delta_Z)^2\Big]\Big[(\omega-\Delta_Z)^2-4 \alpha^2 k^2\Big]}}{
\cosh\left(\frac{k^2}{2mT} -\frac{E_F}{T}\right ) + \cosh \left ( \frac{\omega}{2T} \right )}, \nonumber \\
\end{eqnarray}
\begin{eqnarray}
\label{parallelT}
&&\hspace{-0.4cm}\sigma_{\|}(\omega)=\left ( \frac{e\, \alpha}{2\sqrt{\pi} \omega\Delta_Z} \right )^2
 \sinh \left( \frac{\omega}{2T} \right )
\int_0^\infty dk \,\,k
 \nonumber \\
&&\times\frac{(4\alpha^2 k^2 - \Delta_Z^2 -\omega^2)^2}
{\sqrt{\Big[4 \alpha^2 k^2 - ( \omega+ \Delta_Z)^2 \Big]\Big[ ( \omega- \Delta_Z)^2 -4 \alpha^2k^2\Big]}}
\nonumber \\
&&\times\frac{1}
{\cosh \left(\frac{k^2}{2mT} -\frac{E_F}{T} \right)+\cosh\left (\frac{\omega}{2T}\right) }\!\!.
\nonumber \\
\end{eqnarray}
Expressions, Eqs. (\ref{Tperpabsorp}) and (\ref{parallelT}), describe the broadening of the optical conductivity peak at finite temperature. We can expect on general grounds that the
chiral resonance broadens with $T$ much faster than the EDSR peak. This is because
the distance between the two
Zeeman-split branches of the spectrum is $\Delta_Z$ at any momentum, while the SO-splitting
is proportional to the momentum. In the latter case, the transitions between the states away from the Fermi surface, which become possible with increasing $T$, have different frequencies.
For the same reason, the low-frequency absorption that becomes possible near the condition $\Delta_{SO}=\Delta_Z$ is
most sensitive to increasing $T$.
\begin{figure}[t]
\includegraphics[width=80mm]{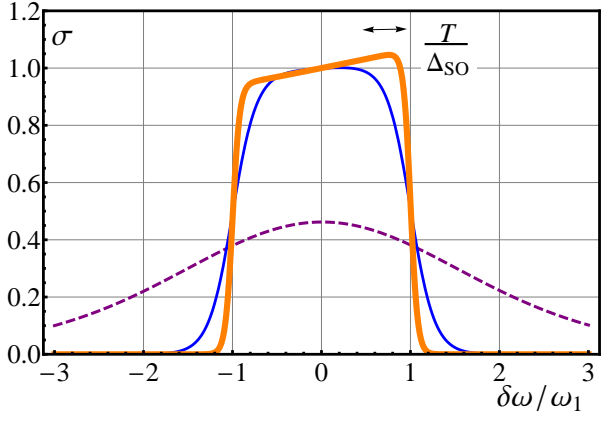}
\caption{The lineshape of the chiral resonance is plotted from Eq. (\ref{chiralab})
in the units of $e^2/16 $ for three dimensionless temperatures,  $\Delta_{SO}/T$,
and spin-orbit coupling strength $\Delta_{SO}/E_F=0.01$. In the low-temperature limit,  $\Delta_{SO}/T=50$ (thick line), the lineshape is box-like.
At temperature $\Delta_{SO}/T=15$ (thin line) the edges of the box get smeared.
Complete smearing corresponding to $\Delta_{SO}/T=2$ is illustrated with dashed line.}
\label{spinonly}
\end{figure}
\subsubsection{Temperature smearing of the chiral resonance}
In particular case of the chiral resonance, $\Delta_Z \rightarrow 0$, the integration over $k$ can be performed explicitly, since the integrand is non-zero only in the vicinity of $k=\omega/2 \alpha$. Then it is sufficient to substitute $k=\omega/2 \alpha$ into the denominator. This leads\cite{Magarill} to the following finite-$T$ generalization of Eq. (\ref{Farid1})
\begin{eqnarray}
\label{chiralab}
\sigma(\delta\omega)\!\!\!&=&\!\!\! \left ( \frac{e^2}{32} \right )
\nonumber \\
&&\hspace{-0.2cm}\times\left (\frac{\sinh\left(\frac{\Delta_{SO}+\delta\omega}{2T}\right)}
{\cosh\left [ \frac{\Delta_{SO}}{4T} \left ( 1 - \frac{\delta \omega}{\omega_1}\right ) \right ]
\cosh\left [ \frac{\Delta_{SO}}{4T} \left ( 1 + \frac{\delta \omega}{\omega_1}\right ) \right ]}\right )\!\!. \nonumber \\
\end{eqnarray}
At low $T\ll \Delta_{SO}$ the hyperbolic sine and cosines can be replaced by the corresponding
exponents. Then we find that, outside the plateau, $\sigma(\omega)$ falls off as $\exp (-\frac{\Delta_{SO}}{2T\omega_1}|\delta\omega\pm \omega_1|)$. Thus, the plateau is
only slightly rounded near the edges, namely, within the interval $|\delta \omega\pm \omega_1 | \sim T\omega_1/ \Delta_{SO} \ll \omega_1$.
Upon increasing $T$, at $T \gg \Delta_{SO}$, we can neglect $1$ in both $\cosh$-terms. Then the denominator will become $\cosh^2(\Delta_{SO} \delta\omega/4T\omega_1)$. We see that characteristic width of the absorption spectrum is $\sim T\omega_1/\Delta_{SO}$, which is much bigger than $\omega_1$. Note however, that, since $\omega_1=\Delta_{SO}^2/4E_F$, this width remains smaller than $\Delta_{SO}$ as long as $T$ remains smaller than $E_F$.
Evolution of the shape of the chiral resonance with $T$ is illustrated in Fig.~\ref{spinonly}.
\subsubsection{Temperature smearing of EDSR }
Consider now the opposite limit of EDSR. The prime modification which finite but low temperature
imposes on $T=0$ expression Eq. (\ref{EDSRperp}) is a smearing of the square-root singularities
at the edges. Due to temperature smearing the factor $(|\omega-\Delta_Z| \pm \Delta_{SO})^{1/2}$
gets replaced by $\left(\omega_1T/\Delta_{SO}\right)^{1/2}$.
To see this, it is convenient to transform from integration over $k$ to integration over $r$, as it was done at $T=0$. Then from Eq. (\ref{Tperpabsorp}) we get
\begin{eqnarray}
\label{TempEDSR}
&&\sigma_\perp = \left ( \frac{e^2}{8 \pi} \right )\int_{-1}^1 dr \sqrt{ 1- r^2}\nonumber \\
&&\times\frac{\sinh(\omega/2T)}
{\cosh\left(\frac{2\Delta_Z \omega r + \Delta_Z^2 +\omega^2 -\Delta_{SO}^2}{4\omega_1 T}\right) + \cosh(\omega/2T)}.
\end{eqnarray}
It is seen that for $T \ll \Delta_Z$ we reproduce the zero-temperature result. In the opposite limit, $T \gg \Delta_Z$, the major contribution to the integral still comes from $r$ close to $-1$. In this limit, the hyperbolic sine and cosine can be replaced by the argument and $1$, respectively.  Introducing the dimensionless variables, $v=(\Delta_Z \omega/2\omega_1 T)(r+1)$ and
\begin{equation}
a(\omega)= \frac{
(\Delta_Z -\omega)^2-\Delta_{SO}^2 }{2\Delta_Z \omega},
\end{equation}
we can rewrite $\sigma_\perp$ as
\begin{equation}
\label{highTperp}
\sigma_\perp(\omega)= \left(\frac{e^2}{4\pi}\right)
\left(\frac{\omega_1}{\Delta_Z}\right)^{3/2}
\left(\frac{T}{\Delta_Z}\right)^{1/2}
g_\perp\left(\frac{|a(\omega)| \Delta_Z \omega}{2\omega_1T}\right),
\end{equation}
where the dimensionless function $g_\perp(x)$ is defined as
\begin{equation}
\label{fperp}
g_\perp(x)=\int_{0}^\infty \frac{dv \sqrt{v}}{\cosh(x \pm v) +1}.
\end{equation}
In the above definition the sign is minus for $a(\omega)>0$ and plus for $a(\omega)<0$. Consider first positive $a(\omega)$.
For $x \gg 1$, the function $g_\perp(x)$ behaves as $2x^{1/2}$, and we reproduce the zero-temperature result.
For $x \ll 1$, we can replace $g_\perp(x)$ by $g(0)\approx 1.07$,
so that
\begin{equation}
\label{asymptote}
\sigma_\perp= \left(\frac{e^2}{4\pi}\right)
\left(\frac{\omega_1}{\Delta_Z}\right)^{3/2}
\left(\frac{T}{\Delta_Z}\right)^{1/2}\hspace{-0.5cm}.
\end{equation}
Comparing Eqs. (\ref{fperp}) and (\ref{EDSRperp}), we conclude that the smearing of
the square-root singularity takes place within the interval, $(\Delta_Z-\omega-\Delta_{SO}) \sim \omega_1 T/ \Delta_{SO}$.

The large-$x$ asymptote, for negative $a$, of $g_\perp$ is $g_\perp= \sqrt{\pi}e^{-x}$. The exponential describes the decay of $\sigma_\perp$ in the region of $\omega$ where it was zero at $T=0$.

In a similar way, the smearing of inverse square-root divergence in Eq. (\ref{EDSRparallel}) is described by the function $g_\|(x)$ defined as
\begin{equation}
\label{fpara}
g_\|(x)=\int_{0}^\infty \frac{dv}{ \sqrt{v}\big(\cosh(x \pm v) +1)}.
\end{equation}
Then Eq. (\ref{parallelT}) reads
\begin{equation}
\label{limit high T perp}
\sigma_\|(\omega)= \left ( \frac{e^2}{16\pi} \right )  \left(\frac{\omega_1}{T}\right )^{1/2}
g_\|\left(\frac{|a(\omega)| \Delta_Z \omega}{2\omega_1T}\right).
\end{equation}
For positive $a(\omega)$ in the limit $x\gg1$, the function $g_\|$ tends to $2/\sqrt{x}$ and Eq. (\ref{limit high T perp}) reduces to Eq. (\ref{EDSRparallel}).
The inverse square-root singularity is smeared in the same domain, $(\Delta_Z-\omega-\Delta_{SO}) \sim \omega_1 T/ \Delta_{SO}$, as for $\sigma_{\perp}$, where the argument of  $g_\|$ is $\sim 1$. In this domain the ratio $\sigma_\|/\sigma_\perp$ remains big but finite, $\sigma_\|/\sigma_\perp\sim \frac{\Delta_Z^2}{\omega_1 T}$.
For $a(\omega)$ negative, in the large-$x$ limit, we have $g_\|=2\sqrt{\pi}e^{-x}$. Substituting this $g_\|$ into Eq. (\ref{limit high T perp}), we find that to the left of the smeared anomaly $\sigma_\|$ decays as
\begin{equation}
\sigma_\|=\left(\frac{e^2}{8 \sqrt{\pi}}\right)
\left(\frac{\omega_1}{T}\right)^{1/2}
\exp\left[ \frac{-\Delta_{SO}}{2\omega_1  T}\big | \Delta_{SO} -( \omega -\Delta_Z)\big|\right].
\end{equation}
\subsubsection{$\Delta_Z =\Delta_{SO}$ at finite $T$}
As stated in the Introduction, the most nontrivaial situation corresponds to $\Delta_Z= \Delta_{SO}$, where the two Fermi surfaces undergo restructuring.
The expressions for $T=0$ optical conductivity under this condition
read
\begin{equation}
\label{perp1T0}
\sigma_{\perp}(\omega)=\left ( \frac{e^2}{8 \pi} \right )\frac{\omega_1}{\Delta_Z^2} \sqrt{ 4\Delta_Z^2- \omega^2 },
\end{equation}
\begin{equation}
\label{parallel1T0}
\sigma_{\|}(\omega)
=\left ( \frac{e^2}{8 \pi} \right ) \frac{\omega_1}{\Delta_Z^2}
\frac{\omega^2}{\sqrt{ 4\Delta_Z^2- \omega^2}},
\end{equation}
We will now analyze how temperature affects the low-$\omega$ part of absorption and the sharp edge at $\omega=2\Delta_Z$.

It is convenient to make
the change of variable $k^2= 2m(E_F + QT)$ in Eq. (\ref{Tperpabsorp}), after which it takes the form
\begin{figure}[t]
\includegraphics[width=70mm]{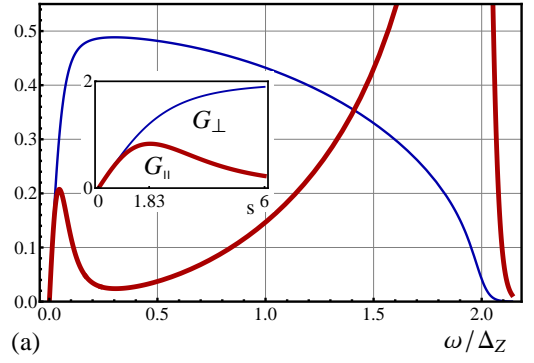}
\hspace{.5 cm}
\includegraphics[width=70mm]{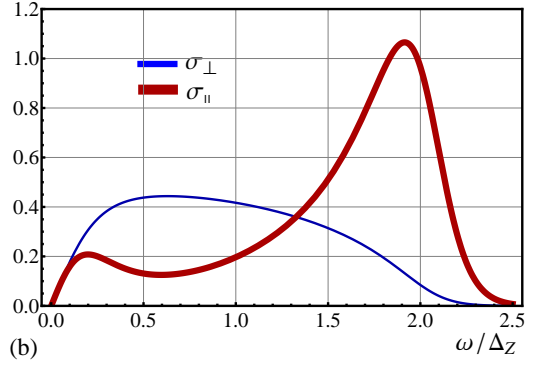}
\hspace{.5cm}
\includegraphics[width=70mm]{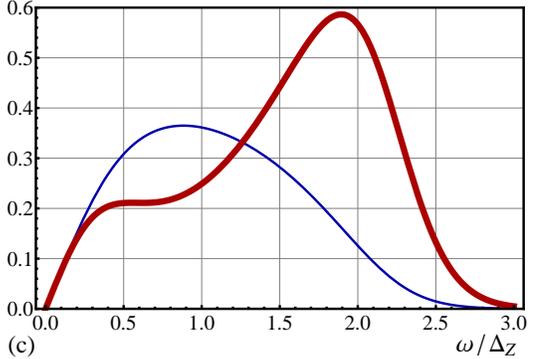}
\caption{
 Evolution of optical conductivity with $T$ in the regime $\Delta_Z=\Delta_{SO}$.
 Eqs. (\ref{sigmatpar}) and (\ref{sigmaytperp}), are plotted in the units $(e^2/16 \pi)(\Delta_Z/ E_F)$ for three values of $T/E_F$; (a) low temperature limit $T/E_F=0.05$. (b) intermediate  temperature $T/E_F=0.2$. (c) ``high" temperature $T/E_F=0.5$. Inset in (a) shows the dimensionless functions $G_{\perp}(s)$ and $G_{\|}(s)$ calculated from Eqs. (\ref{Gperp}) and (\ref{Gpara}), respectively.}
\label{zeeman2}
\end{figure}
\begin{eqnarray}
\label{perpendicular5}
&&\hspace{-0.4cm}\sigma_{\perp}(\omega)=
\left( \frac{e^2}{32 \pi \omega^2 } \right )\! \left (\frac{\Delta_{SO}}{\Delta_Z} \right)^2\!\! \left ( \frac{ T}{E_F} \right)
\sinh \left (\frac{\omega}{2T} \right )
\! \int \!dQ
\nonumber \\
&&\hspace{-0.1cm}\times\frac{
\sqrt{ 4 \Delta_{SO}^2 \Delta_Z^2 \left( 1 + \frac{Q T}{E_F}\right)\! - \!\bigg( \Delta_{SO}^2 \left(1 +\frac{ QT}{E_F}\right)\! + \Delta_Z^2 -\omega^2 \bigg)^2}
}{\cosh Q +\cosh\left (\frac{\omega}{2T}\right)}.\nonumber \\
\end{eqnarray}
Now look at the point where
$\Delta_{SO}= \Delta_Z$, in the limit $T \gg \omega$. Then Eq. (\ref{perpendicular5})
reduces to
\begin{equation}
\label{sigmatpar}
\sigma_\perp(\omega)\!=\!
\left ( \frac{e^2}{64 \pi } \right ) \!
 \left ( \frac{\Delta_Z^2}{ \omega E_F} \right )
{\int} dQ
\frac{\sqrt{\frac{4\omega^2 }{\Delta_Z^2}-\left ( \frac{QT}{E_F}-\frac{\omega^2}{\Delta_Z^2}\right)^2}}{\cosh Q +1}.
\end{equation}
Similarly, the $T\gg \omega$ limit of Eq. (\ref{parallelT}) assumes the form
\begin{eqnarray}
\label{sigmaytperp}
\hspace{-0.6cm}\sigma_\|&&\hspace{-0.4cm}(\omega)\!=\!
\left ( \frac{e^2}{64 \pi } \right ) \!
 \left ( \frac{\Delta_Z^2}{ \omega E_F} \right )
\nonumber\\
&&\times{\int} dQ
\frac{\left ( \frac{QT}{E_F}-\frac{\omega^2}{\Delta_Z^2}\right)^2}{\big(\cosh Q +1\big)\sqrt{\frac{4\omega^2 }{\Delta_Z^2}-\left ( \frac{QT}{E_F}-\frac{\omega^2}{\Delta_Z^2}\right)^2}}.
\end{eqnarray}
Consider first the domain $\omega \ll \Delta_Z$.
Then in the combination $( \frac{QT}{E_F} - \frac{\omega^2}{\Delta_Z^2})$ in the integrands of Eqs. (\ref{sigmatpar}) and (\ref{sigmaytperp}), the term $(\omega/\Delta_Z)^2$ can be dropped. This allows us to simplify Eqs. (\ref{sigmatpar}) and (\ref{sigmaytperp}) to
\begin{equation}
\sigma_\perp(\omega)\!=\!
\left ( \frac{e^2}{32 \pi } \right ) \!
 \left ( \frac{\Delta_Z}{ E_F} \right ) G_\perp\left( \frac{2\omega E_F}{\Delta_ZT}\right),
\end{equation}

\begin{equation}
\sigma_\|(\omega)\!=\!
\left ( \frac{e^2}{32 \pi } \right ) \!
 \left ( \frac{\Delta_Z}{ E_F} \right ) G_\|\left( \frac{2\omega E_F}{\Delta_ZT}\right),
\end{equation}
where the functions $G_\perp$ and $G_\|$ are defined as
\begin{equation}
\label{Gperp}
G_\perp(s)= \frac{1}{s}\int_{-s}^s \frac{dQ}{\cosh Q +1} \sqrt{s^2-Q^2},
\end{equation}
\begin{equation}
\label{Gpara}
G_\|(s)= \frac{1}{s}\int_{-s}^s \frac{dQ}{\cosh Q +1}\frac{Q^2}{ \sqrt{s^2-Q^2}}.
\end{equation}
These functions are plotted in Fig.~\ref{zeeman2}a inset. At $s \ll 1$ they both behave as
$\pi s/4$. For $s \gg 1$ the function $G_\perp$ saturates at $G_\perp(\infty)=2$, while the function $G_\|$ falls off as $G_\|(s) \approx 2\pi^2/3s^2$.

It can be seen from Fig.~\ref{zeeman2}a inset that $G_\|(s)$ has a maximum at $s=\tilde{s}=1.83$. This translates into the maximum in absorption at frequency
\begin{equation}
\tilde{\omega}=\frac{0.91\Delta_Z T}{E_F}.
\end{equation}
This sharp maximum clearly shows in the absorption lines plotted in Fig.~\ref{zeeman2}a,b which were calculated numerically from the full expression Eq. (\ref{sigmaytperp}) for particular temperatures, $T/E_F=0.05$ and $T/E_F=0.2$. Remarkably, the position of maximum moves linearly with temperature for low temperatures. The maximum starts to disappear when temperature becomes high enough, $T/E_F=0.5$.

It is also seen from Figs.~\ref{zeeman2}b,c that the right boundary $\omega \approx 2 \Delta_Z$ of the absorption spectrum is smeared with temperature. This smearing can be described analytically in terms of the functions $g_\perp$ and $g_\|$ defined by Eqs. (\ref{fperp}) and (\ref{fpara}). Indeed, for $\omega$ close to $2 \Delta_Z$, the square root in the numerator of Eq. (\ref{sigmatpar}) can be simplified to $\sqrt{2}(\omega/\Delta_Z) \big[ \frac{QT}{E_F} - \frac{2}{\Delta_Z}(\omega-2\Delta_Z)\big]^{1/2}$. Then $\sigma_\|(\omega)$ and $\sigma_\perp(\omega)$ acquire the form
\begin{equation}
\sigma_\perp(\omega)= \!\left(\! \frac{\sqrt{2}e^2}{64 \pi}\!\right)\!\!
\left(\frac{\Delta_Z}{E_F}\right)\!
\left(\frac{T}{E_F}\right)^{1/2}\!\!
g_\perp\left(\!\frac{2E_F}{T}\frac{(\omega-2\Delta_Z)}{\Delta_Z}\!\right)\!,
\end{equation}
\begin{equation}
\sigma_\|(\omega)= \!\left(\! \frac{e^2}{64\sqrt{2} \pi}\!\right)\!\!
\left(\frac{\Delta_Z}{E_F}\right)\!
\left(\frac{E_F}{T}\right)^{1/2}\!\!
g_\|\left(\!\frac{2E_F}{T}\frac{(\omega-2\Delta_Z)}{\Delta_Z}\!\right)\!.
\end{equation}
It is worth noting that even at high temperature $T=0.5E_F$  the right boundary $\omega=2\Delta_Z$ is very well pronounced.

\section{ESR lineshape}
\label{sec:magnetic}
In this section we use the
above results for the absorption of the electric
field to calculate the ESR lineshape.

For ESR, the corresponding
interaction Hamiltonian of spins with the ac magnetic field reads

\begin{equation}
\hat{H}_{int}^m=\gamma \big(2 \hat{\mathbf{s}}\cdot\mathbf{B}\big) \cos\,\omega t,
\end{equation}
where $\gamma$ is the gyromagnetic ratio.
The quantity describing the ESR strength, imaginary part of susceptibility, can be expressed via the eigenfunctions of the free Hamiltonian
Eq. (\ref{Hamiltonian EDSR}) as follows
\begin{eqnarray}
\label{main2}
\chi_i(\omega)\!\!\!&=&\!\!\! 8\pi \gamma^2\big(1-e^{-\omega/T}\big )\sum_{\bf k}  \big|\langle u_\mathbf{k}^{(1)}\big| \hat{s}_i \big |u_{\mathbf{k}}^{(2)} \rangle\big|^2
\nonumber \\
 &&
\times \delta\big( \mathcal{E}_\mathbf{k}^{(2)} - \mathcal{E}_\mathbf{k}^{(1)} - \omega \big)\!\Big[1- f\big( \mathcal{E}_\mathbf{k}^{(2)} \big)\!\Big]
 f\big(\mathcal{E}_\mathbf{k}^{(1)}\big),\nonumber \\
\end{eqnarray}
where $i$ is the polarization axis of the ac magnetic field.

Our prime observation that the values $\chi_i$ can be expressed in a simple
way though the components of optical conductivity, $\sigma_{\perp}(\omega)$ and $\sigma_{\|}(\omega)$, calculated above.
Consider, for example, ac magnetic field polarized along $x$. Then, using Eqs. (\ref{spectrum}), (\ref{spinor}), the matrix element
$\langle u_\mathbf{k}^{(2)}\big| \hat{s}_x \big |u_{\mathbf{k}}^{(1)} \rangle$ can be expressed
as
\begin{equation}
\label{sx}
\langle u_\mathbf{k}^{(1)}\big| \hat{s}_x\big |u_{\mathbf{k}}^{(2)} \rangle
=i\frac{\alpha k_x - \Delta_Z/2}{\mathcal{E}_\mathbf{k}^{(2)}-\mathcal{E}_\mathbf{k}^{(1)}}.
\end{equation}
Upon transformation to polar coordinates, the numerator of Eq. (\ref{sx}) turns into $k\cos\phi - \Delta_Z/2$,
while the $\delta$-function in Eq. (\ref{main2}) insures that the denominator is equal to $\omega$.
In this way the summation over momenta in Eq. (\ref{main2}) reduces to same integration as in Eq. \eqref{parallel} and we arrive at the relation
\begin{equation}
\label{relation}
\chi_x(\omega)=\frac{2\gamma^2}{e^2\alpha^2}\omega\sigma_{\|}(\omega).
\end{equation}
From comparing Eq. (\ref{main2}) to Eq. (\ref{main}), it is obvious that this relation applies at all temperatures.

Similarly, the matrix element $\langle u_\mathbf{k}^{(1)}\big| \hat{s}_y\big |u_{\mathbf{k}}^{(2)} \rangle$ in polar coordinates assumes the form
 $  i \alpha k \sin\phi/(\mathcal{E}_\mathbf{k}^{(2)}-\mathcal{E}_\mathbf{k}^{(1)})$.
This makes Eq. \eqref{main2} for $i=y$ proportional to $\sigma_{\perp}$ in Eq. \eqref{zeemanab}, and leads to
\begin{equation}
\label{relation1}
\chi_y(\omega)=\frac{2\gamma^2}{e^2\alpha^2}\omega\sigma_{\perp}(\omega).
\end{equation}
Concerning polarization of magnetic field along $z$,
we have
\begin{equation}
\label{sz}
\langle u_\mathbf{k}^{(1)}\big| \hat{s}_z\big |u_{\mathbf{k}}^{(2)} \rangle
=-1,
\end{equation}
so that summation over momenta does not contain angular dependence at all. This allows to
relate $\chi_z(\omega)$ to the sum of $\sigma_{\perp}(\omega)$ and $\sigma_{\|}(\omega)$ as follows
\begin{equation}
\label{final}
\chi_z= \frac{2\gamma^2}{e^2\alpha^2}\omega
\big[\sigma_\perp(\omega)+\sigma_\|(\omega)\big].
\end{equation}
We see that ESR line shape for different polarizations of the ac magnetic field can be written
in terms of optical conductivity $\sigma_{\perp}$ and $\sigma_{\parallel}$. Response to the ac field polarized perpendicular
to the static field, namely,  $\chi_{x}$ and $\chi_{z}$, is determined mostly by $\sigma_{\parallel}(\omega)$
which diverges in an inverse square-root fashion at the boundaries of the absorption interval. In weak static field the lineshape
is a $\delta$-peak at $\omega=\Delta_{SO}$\cite{Loss}.

It is natural that the ac field parallel to the static field
causes less singular response, than the ac field normal to the
static field. Indeed,  $\chi_y$ is proportional to $\sigma_{\perp}(\omega)$, which exhibits square-root suppression
at the boundaries of the absorption interval.

\section{Discussion}

Interplay of spin-orbit interaction and in-plane magnetic field results in strongly
anisotropic line shape of ESR response of the two-dimensional fermion gas.
The response is most singular at the upper and lower boundaries of fermion
continuum when the ac magnetic field is perpendicular to the static
 magnetic field, see Fig.~\ref{figzeeman1} and Section \ref{sec:magnetic}. The behavior of the ESR lineshape is
 most ``lively" near the ``compensation" point $\Delta_Z=\Delta_{SO}$,
 where it exhibits a nontrivial temperature dependence up to
 high temperature $T\sim E_F$.

A possible way to reveal these features experimentally is to
fix the polarization of microwave radiation and to rotate the sample with respect to the external static magnetic field.

Restricting two-dimensional gas to a quantum wire geometry when the motion
along, say, $y$-axis is quantized, will result in ``slicing'' the Fermi-surface
along several allowed $k_y$ momenta \cite{rachel_thesis}. In the extreme case of the single
channel quantum wire, when only the lowest transverse subband is populated,
the ESR lineshape will be reduced to two delta-function like features at
the lower and upper boundaries of the ESR continuum. This is exactly
the response found previously\cite{suhas2008}
within a different, purely one-dimensional
treatment of an insulating Heisenberg spin chain with uniform DM
interaction.
Ref. \onlinecite{artem} finds a very similar response
of a conducting quantum wire with spin-orbit interaction to ac electric field.
The correspondence between the two above mentioned one-dimensional studies illustrates nicely the central point of our work that ESR is
an effective probe of the Fermi surfaces irrespective of their origin.
 Both conventional Fermi surface\cite{artem} and spinon Fermi surface\cite{suhas2008} give rise to the same ESR response.

One-dimensional geometry  also allows one
to address effects of electron-electron interactions  omitted in the present work.
Although relatively marginal even in one-dimensional geometry \cite{suhas2008},
these are difficult to account for at finite frequencies as recent detailed investigation
shows \cite{karimi2011}. Extending this approach to two-dimensional systems,
along the lines of renormalization group calculations in \cite{zak2010}, represents
an interesting outstanding problem.

Throughout the paper we neglected the disorder. In realistic sample, with finite disorder, the absorption peak calculated above will always reside on the
background of the Drude peak coming from the intraband scattering.
The condition that disorder does not smear the interband absorption peak
reads: $\Delta_{Z}\tau, \Delta_{SO}\tau \gg 1$, where $\tau$ is the scattering
time. Under this condition, the peak  which we calculated is distinguishable,
since the background is relatively flat. However, in order for interband
peak to dominate over the intraband contribution, a more stringent condition
must be met. Namely, the products $\Delta_{Z}\tau, \Delta_{SO}\tau$ must exceed
$(E_F\tau)^{1/2}$.

It is worth noting that ideas outlined here have been recently successfully
tested experimentally \cite{povarov} in quasi-one-dimensional spin-1/2 antiferromagnet Cs$_2$CuCl$_4$
where uniform DM interaction on chain bonds was found to split the ESR signal into two peaks already
in the finite-temperature spin-liquid phase, just as we describe above. Many higher-dimensional candidate
spin-liquid materials, such as two-dimensional $\kappa$-(BEDT-TTF)$_2$Cu$_2$(CN)$_3$ \cite{2d_organic} and three-dimensional
Na$_4$Ir$_3$O$_8$ \cite{3d_spinliquid}, have been argued to possess spin-liquid ground states with spinon Fermi surfaces
\cite{lee2005, motrunich2005, lawler2008,podolsky2009, pesin2010}.
DM and other kinds of spin-orbit interactions are most certain to be present in these materials --
this is certainly the case for Na$_4$Ir$_3$O$_8$ and highly likely for the organic material as well
(its close relative, $\kappa$-(BEDT-TTF)$_2$Cu[N(CN)$_2$]Cl,
does have substantial DM interaction \cite{2d_organic_so}).
Our work shows that spin-orbit interaction can turn ESR absorption experiment into a insightful probe
of the unusual spinon Fermi surfaces in these and related materials.

\begin{acknowledgements}
We would like to thank L. Balents and E. G. Mishchenko for useful discussions.
This work is supported by NSF awards DMR-0808842 (R.G. and O.S.)  and MRSEC DMR-1121252 (M.R.).
\end{acknowledgements}

\end{document}